\documentclass[a4paper]{jpconf}
\usepackage{graphicx}
\begin{document}

\title{A global DGLAP analysis of nuclear PDFs}

\author{K. J. Eskola$^a$, V. J. Kolhinen$^a$, 
H. Paukkunen$^{a}$\footnote{The speaker} and C. A. Salgado$^b$}

\address{$^a$Department of Physics, University of Jyv\"askyl\"a and Helsinki Institute of Physics, Finland}

\address{$^b$Dipartimento di Fisica, Universit\`a  di Roma ``La Sapienza'' and INFN, 
Roma, Italy}

\ead{kari.eskola@phys.jyu.fi, vesa.kolhinen@phys.jyu.fi, \\ \quad \qquad hannu.paukkunen@phys.jyu.fi, carlos.salgado@cern.ch}

\begin{abstract}
In this talk, we shortly report results from our recent global DGLAP analysis of nuclear parton distributions. This is an extension of our former EKS98-analysis improved with an automated $\chi^2$ minimization procedure and uncertainty estimates. Although our new analysis show no significant deviation from EKS98, a sign of a significantly stronger gluon shadowing could be seen in the RHIC BRAHMS data.

\end{abstract}

\section{Introduction}

The global analysis of nuclear parton distributions (nPDFs) is driven by the experimental fact that the deep inelastic 
structure functions $F_2(x,Q^2)$ measured from nuclear targets show a significant deviation from the free proton ones \cite{Arneodo:1992wf,Armesto:2006ph}.

Perhaps the most simple theoretical approach to this observation is to make use of the factorization theorem of QCD that has proven to provide excellent description of inclusive cross-sections in free-nucleon collisions.  In this approach, the cross-sections are of the generic form
\begin{equation} \label{eq:facQCD}
\sigma_{AB\rightarrow h+X} = \sum_{ij} 
f_i^A (x_1,Q^2) \otimes f_j^B(x_2,Q^2) \otimes \sigma^{i+j \rightarrow h+X}, 
\end{equation}
where where $\sigma^{i+j \rightarrow h+X}$ is the perturbative QCD (pQCD) matrix element squared and $f_i$s are the non-perturbative parton densities whose scale evolution obeys the DGLAP equations \cite{DGLAP}.

The purpose of the global analysis of nPDFs is to find out whether the observed differences in the structure functions, the nuclear modifications, can consistently be absorbed in to the input parton densities --- in other words, do the nuclear modifications effectively factorize. If they do, the resulting nPDFs are of great practical interest, since they can be used as a input in any process that can be factorized as in eq.~(\ref{eq:facQCD}). In this framework the deep question about the dynamical origin of nuclear modifications is not addressed, on the contrary, one must be as unbiased to any model as possible.

Indeed, three independent groups have shown that this approach works quite well:
\begin{itemize}
\item
EKS98 \cite{Eskola:1998iy,Eskola:1998df} was the first global analysis demonstrating that using the leading-order (LO) pQCD formalism, requiring momentum and baryon number conservation, one can reproduce the data from measurements of deep inelastic lepton-nucleus scattering (DIS) and Drell-Yan dilepton production (DY) in proton-nucleus collisions. The fit was done only by eye.
\item
HKM \cite{Hirai:2001np} \& HKN \cite{Hirai:2004wq} which are LO QCD analyses as well, were the first ones to exploit the automated $\chi^2$-minimization procedure and to estimate nPDF uncertainties.
\item
nDS \cite{deFlorian:2003qf} brought the global analysis of nPDFs to the next-to-leading order level in pQCD.
\end{itemize}

Based on our earlier work, EKS98, we have performed a global reanalysis of nPDFs published in \cite{Eskola:2007my} and reported here. What was new compared to the EKS98 were the automated $\chi^2$-minimization procedure and a first attemp for estimating the uncertainties. Although our main objective was to see whether we can improve the EKS98-fit and study the uncertainties, this was anyway a necessary stepping stone for us before we can extend our analysis to NLO-level.

Motivated by the BRAHMS data \cite{Arsene:2004ux} on inclusive hadron production in D+Au collision, which show a systematic suppression relative to p+p at forward rapidities, we raise an intriguing question about the possibility of having clearly stronger gluon shadowing than what has been hitherto seen in the global DGLAP analyses.

\section{The framework}

We define the PDFs $f_i^A(x,Q^2)$ of bound protons in a nucleus with mass number $A$ as
\begin{equation}
f_{i}^A(x,Q^2) = R_{i}^A(x,Q^2) f_{i}^{\rm CTEQ6L1}(x,Q^2),
\end{equation}  
where $f_{i}^{\rm CTEQ6L1}$ refers to the latest free proton PDFs by the CTEQ collaboration \cite{Pumplin:2002vw}. For the bound neutrons we assume the isospin symmetry $d_{\rm proton}=u_{\rm neutron}$ and vice versa. What we actually parametrize, are the nuclear modifications $R_{i}^A(x,Q^2)$ at the initial scale $Q_0^2=1.69 \, {\rm GeV}^2$. At present, the lack of precision data forces us to consider only three different modifications: $R_{V}^A(x,Q_0^2)$ for all valence quarks, $R_{S}^A(x,Q_0^2)$ for all sea quarks, and $R_{G}^A(x,Q_0^2)$ for gluons.

The fit functions are parametrized in three pieces (c.f. Fig.~\ref{ini}):
\begin{eqnarray}
  &&R_1^A(x) = c_0^A+(c_1^A+c_2^A x)[\exp(-x/x_s^A) 
           - \exp(-x_a^A/x_s^A)],   \qquad \qquad x\le x_a^A \nonumber \\
  &&R_2^A(x) = a_0^A + a_1^A x + a_2^A x^2 + a_3^A x^3, 
             \qquad \qquad \qquad \quad \, \, \; \; \: \qquad x_a^A \le x \le x_e^A\\ \label{R2} 
  &&R_3^A(x) = \frac{b_0^A-b_1^A x}{(1-x)^{\beta^A}},      
             \qquad \qquad \qquad \qquad \qquad \quad \: \: \qquad \qquad \qquad x_e^A \le x \nonumber,
\end{eqnarray}
The first one covers the region from shadowing to anti-shadowing maximum at $x_a^A$, the second comes down to EMC-minimum at $x_e^A$, and the third is for the Fermi-motion part.

The $A$-dependence of the fit parameters is assumed to follow a power law
\begin{equation}
  z_i^A = z_i^{A_{\rm ref}} (\frac{A}{A_{\rm ref}})^{\,p_{z_i}},
  \label{eq:Adependence}
\end{equation}
where we have chosen Carbon ($A_{\rm ref}=12$) as a reference nucleus.

After fixing the continuity of $R_{i}^A$s and their first derivatives at $x_a^A$ and $x_e^A$, one is still left with $42$ free parameters of which the baryon number and momentum conservation eat only $4$ away. This was still too much in order to obtain converging fits, and lots of manual work was needed too see which parameters were the most relevant ones. At the end there was $16$ fit parameters.

\section{Results \& Error analysis}

\begin{figure}[h]
\centering
\begin{minipage}{16pc}
\includegraphics[width=16pc]{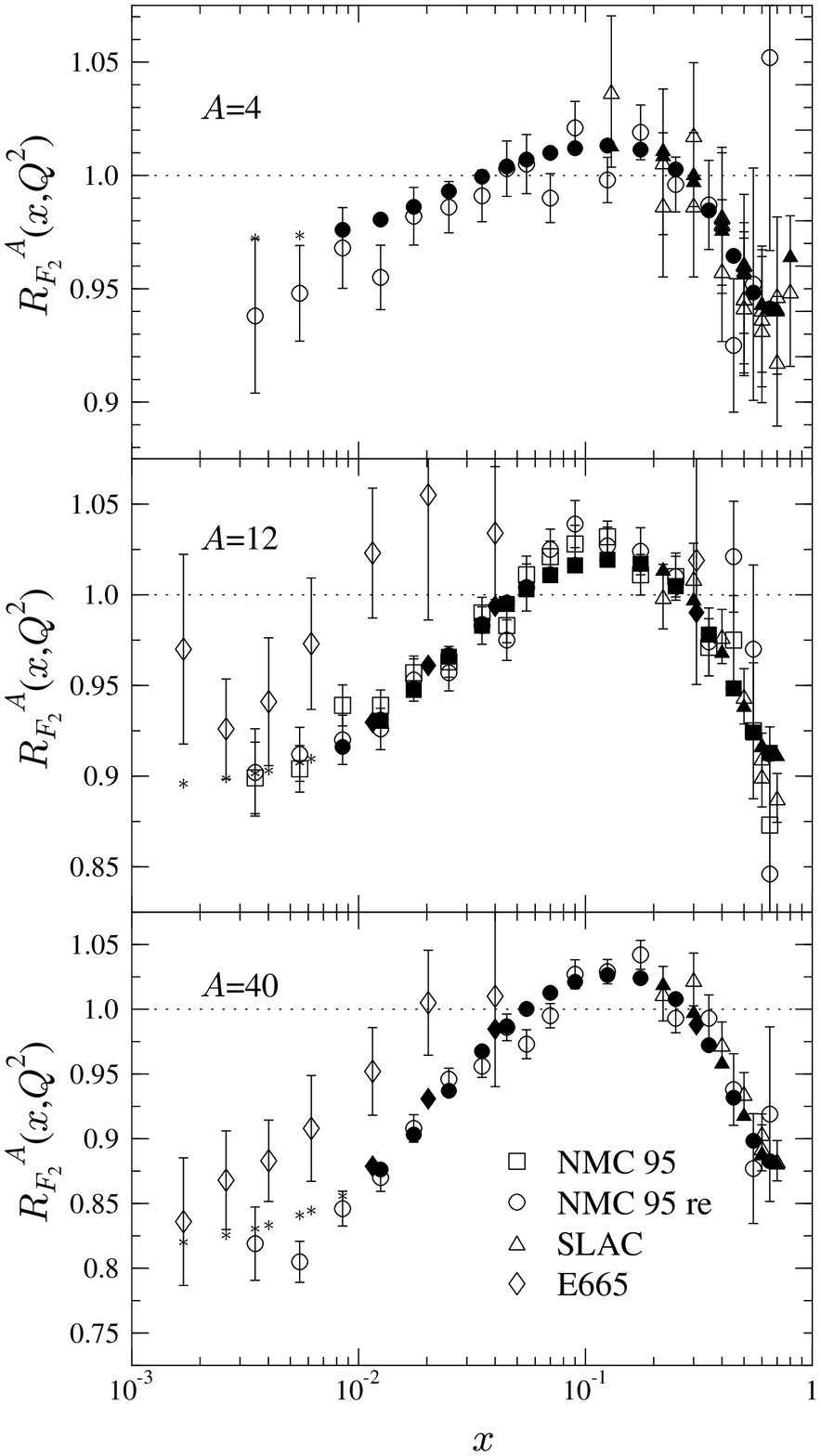}
\end{minipage}\hspace{2pc}%
\begin{minipage}{16pc}
\includegraphics[width=19pc]{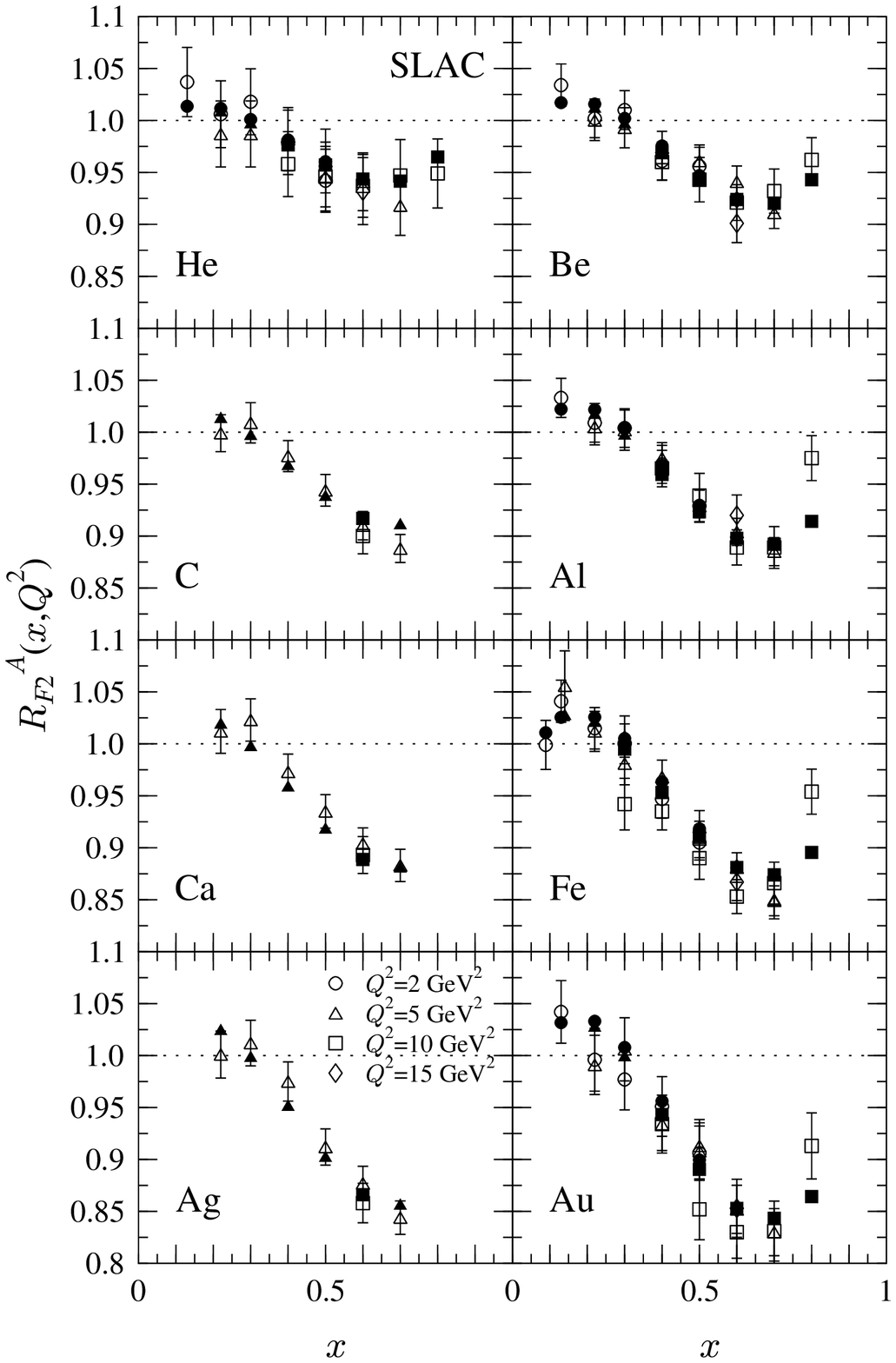}
\end{minipage} 
\caption{\label{fig:data1}Calculated $R_{F_2}^A(x,Q^2)$ (filled symbols) are
  compared to SLAC \cite{Gomez:1993ri}, E665 \cite{Adams:1995is}, NMC 95 \cite{Arneodo:1995cs} 
  and reanalysed NMC 95 data
  \cite{Amaudruz:1995tq}. The asterisks denote our results
  calculated at the initial scale $Q_0^2$, these are for the smallest-$x$ data points 
  whose scales lie in the region $Q^2<Q_0^2$.}
\end{figure}

\begin{figure}[h]
\centering
\begin{minipage}{16pc}
\includegraphics[width=18pc]{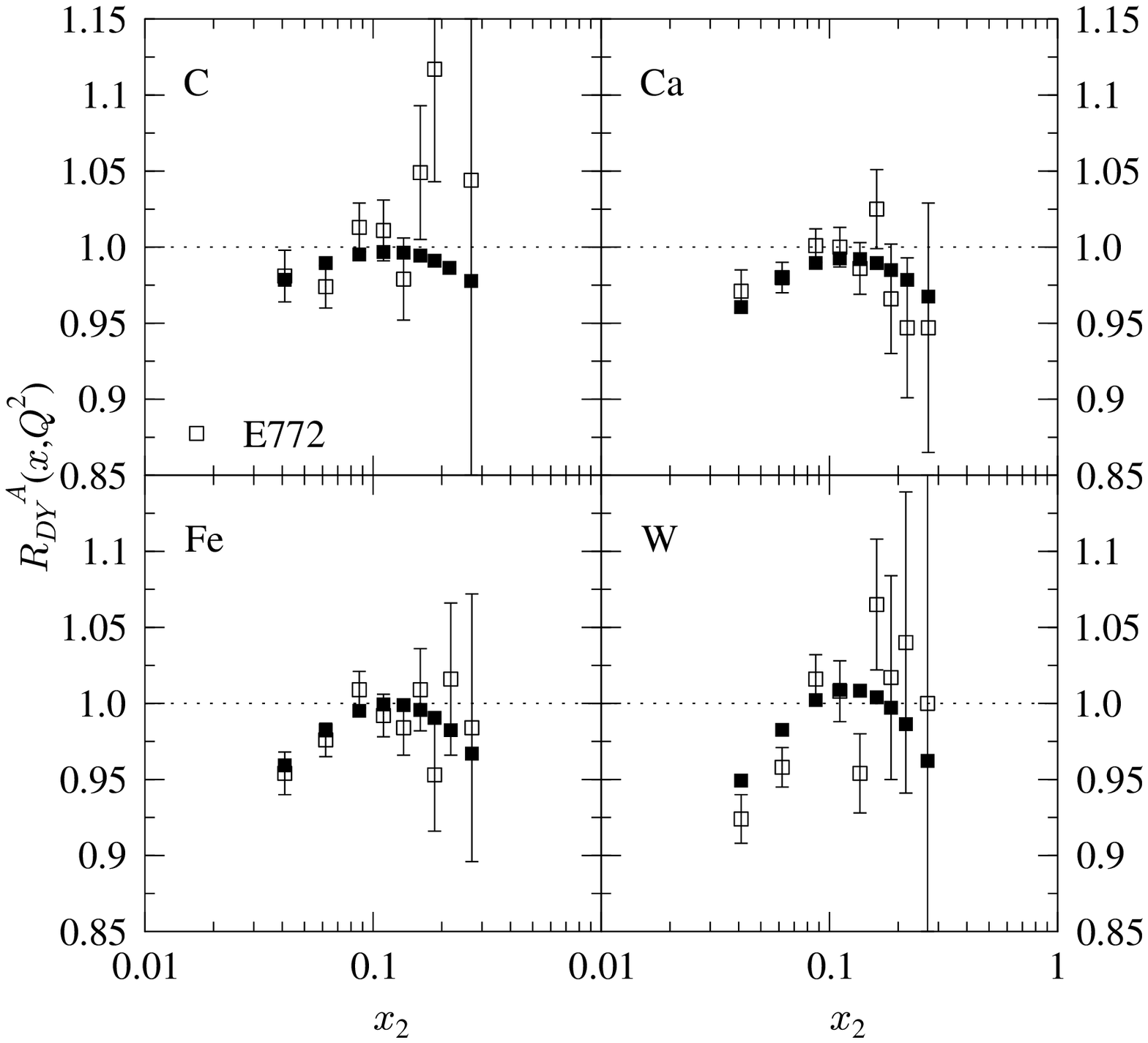}
\end{minipage}\hspace{2pc}%
\begin{minipage}{16pc}
\includegraphics[width=18pc]{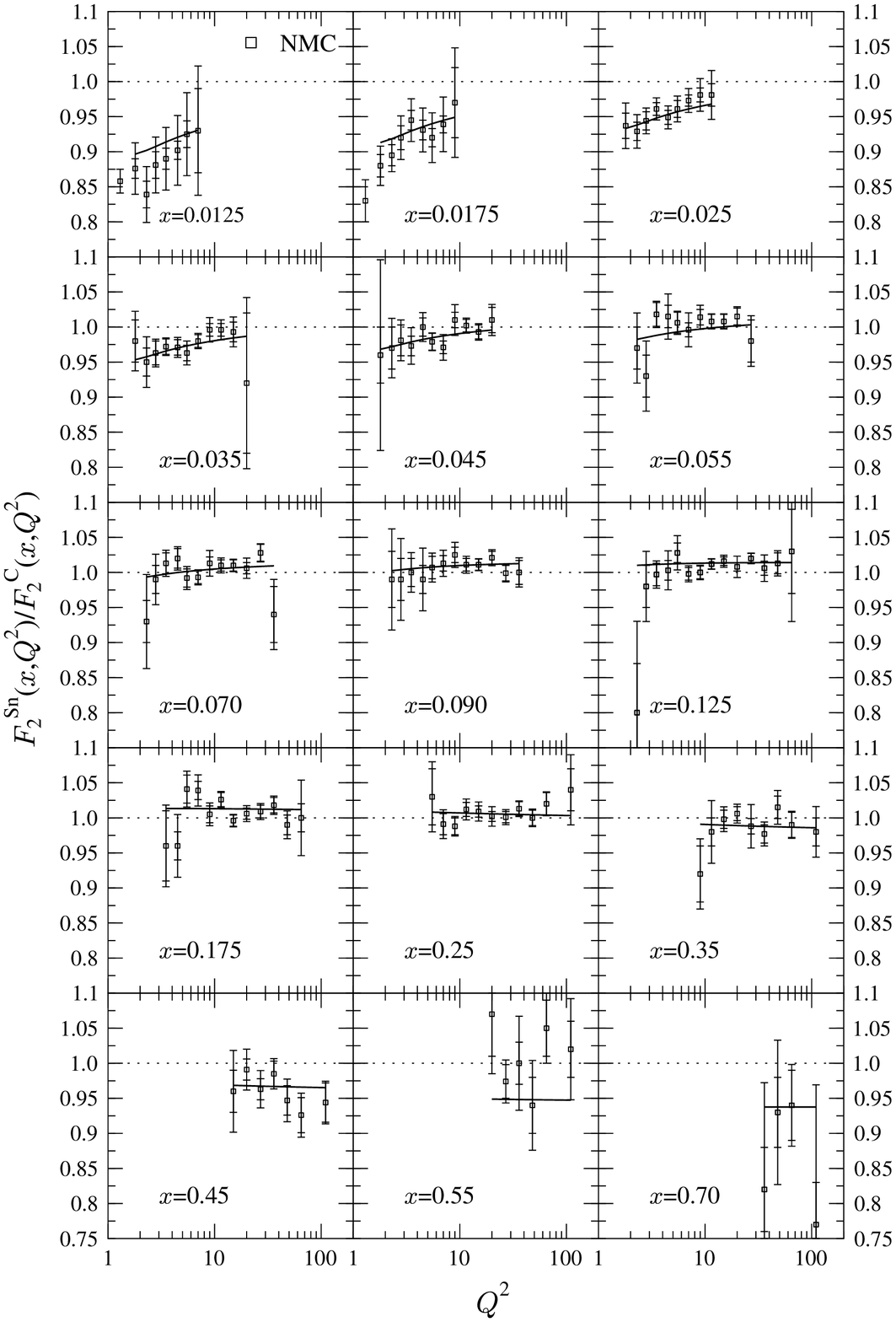}
\end{minipage} 
\caption{\label{fig:data2} Calculated $R_{DY}^A(x,Q^2)$ and $F_2^{\mathrm{Sn}}/F_2^{\mathrm{C}}$ (filled symbols) are compared to E772 \cite{Alde:1990im} and NMC data \cite{Arneodo:1996ru}}
\end{figure}

The experimental input in our analysis was about $500$ points of  DIS- and DY-data in a form
\begin{equation}
 {\rm DIS:} \quad \frac{\frac{1}{A}d\sigma^{lA}/dQ^2dx}{\frac{1}{2}d\sigma^{l{\mathrm D}}/dQ^2dx}
 \,{\buildrel {\rm LO}\over =}\, R_{F_2}^A(x,Q^2),
 \hspace{1cm}
 {\rm DY:}  \quad \frac{\frac{1}{A}d\sigma_{DY}^{{\mathrm p}A}/dx_2 dQ^2}
        {\frac{1}{2}d\sigma_{DY}^{\mathrm {pD}}/dx_2 dQ^2}
		\,{\buildrel {\rm LO}\over =}\,
		R_{DY}^A(x_2,Q^2).
\end{equation}
These covered $11$ different nuclei from Helium up to Lead. Figs. \ref{fig:data1} and \ref{fig:data2} show some of these data and comparison to our result from minimization of
$$\chi^2 = \sum_{i=1}^{N_{\mathrm{data}}} \left( \frac{\mathrm{data}_i-\mathrm 
{theory}_i}{ {\rm error}_i}\right )^2.$$

The number that characterizes the goodness of the fit is $\chi^2/{\rm d.o.f.}$ (${\rm d.o.f} \equiv N_{\rm data} - N_{\rm free \,  parameters}$) which should be less than $1$ if the fit is any good. In our case $\chi^2/{\rm d.o.f.} \sim 0.8$ indicating that the theory fits the data very well and that there is no serious sign why pQCD could not be trusted --- within the considered kinematical range --- also in the nuclear environment. 

\begin{figure}
\centering
 \includegraphics[scale=0.4]{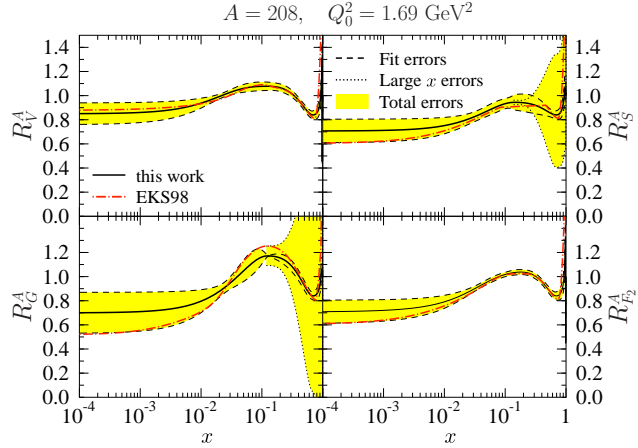}
\caption[]{\small Initial nuclear ratios for Lead together with their uncertainty estimates}  \label{ini}
\end{figure}

The obtained nuclear modifications at the initial scale for Lead are shown in Fig. \ref{ini}.  This figure also presents our uncertainty estimates based on Hessian method of quantifying the uncertainties \cite{Hirai:2003pm}, in which one expands the $\chi^2$  around the minimum w.r.t fit parameters $\xi$ as
\begin{equation}
  \Delta \chi^2 = \chi^2(\hat\xi+\delta\xi) - \chi^2(\hat\xi) 
    = \sum_{i,j} H_{ij} \delta\xi_i \delta\xi_j,
\end{equation}
and the uncertainty of any quantity $F(\hat\xi)$ depending on PDFs is then obtained from
\begin{equation}
  [\delta F(\hat\xi)]^2 = \Delta\chi^2 
   \sum_{i,j} \left(\frac{\partial F(\hat\xi)}{\partial \xi_i}\right)
   H_{ij}^{-1}\left(\frac{\partial F(\hat\xi)}{\partial \xi_j}\right).
  \label{Errors}
\end{equation}
For an ideal $\chi^2$-distribution $\Delta \chi^2 \approx 18$, which we used too, corresponds to ``one sigma''-error. 

Due to technical difficulties obtaining converging fits, the EMC minimum of gluons and sea quarks was fixed to follow the valence quarks at $Q^2_0$. This resulted as an unreliably small error bands for gluons and sea quarks, and they had to be computed separately. They are the ``Large-$x$ errors'' in Fig.~\ref{ini}. The combined uncertainties are shown as a yellow bands. Interestingly, the old EKS98 parametrization lies within these uncertainties and there is no reason to release a new parametrization --- EKS98 works just fine.

However, one should be very cautious about these error bands! First, below $x \sim 10^{-2}$ there is no experimental data above $Q^2=1.69 \, {\rm GeV}^2$, and the behaviour at small-$x$ region is constrained only by the sum rules and is bound to the form of the fit function. Second, the PDFs themselves depend on choices and conventions, like kinematical cuts, choosing the factorization scale, treatment of heavy quarks, and choosing and weighting data sets in forming $\chi^2$. How these choices affect the obtained PDFs is not seen in the error analysis performed here --- the error bands only reflect the experimental data and their errors. Third, there is no well-established way to choose $\Delta \chi^2$, and our choice $\Delta \chi^2 \approx 18$ is probably quite restrictive.

\section{Are the gluons different?}

\begin{figure}[h]
\centering
\begin{minipage}{16pc}
\includegraphics[width=20pc]{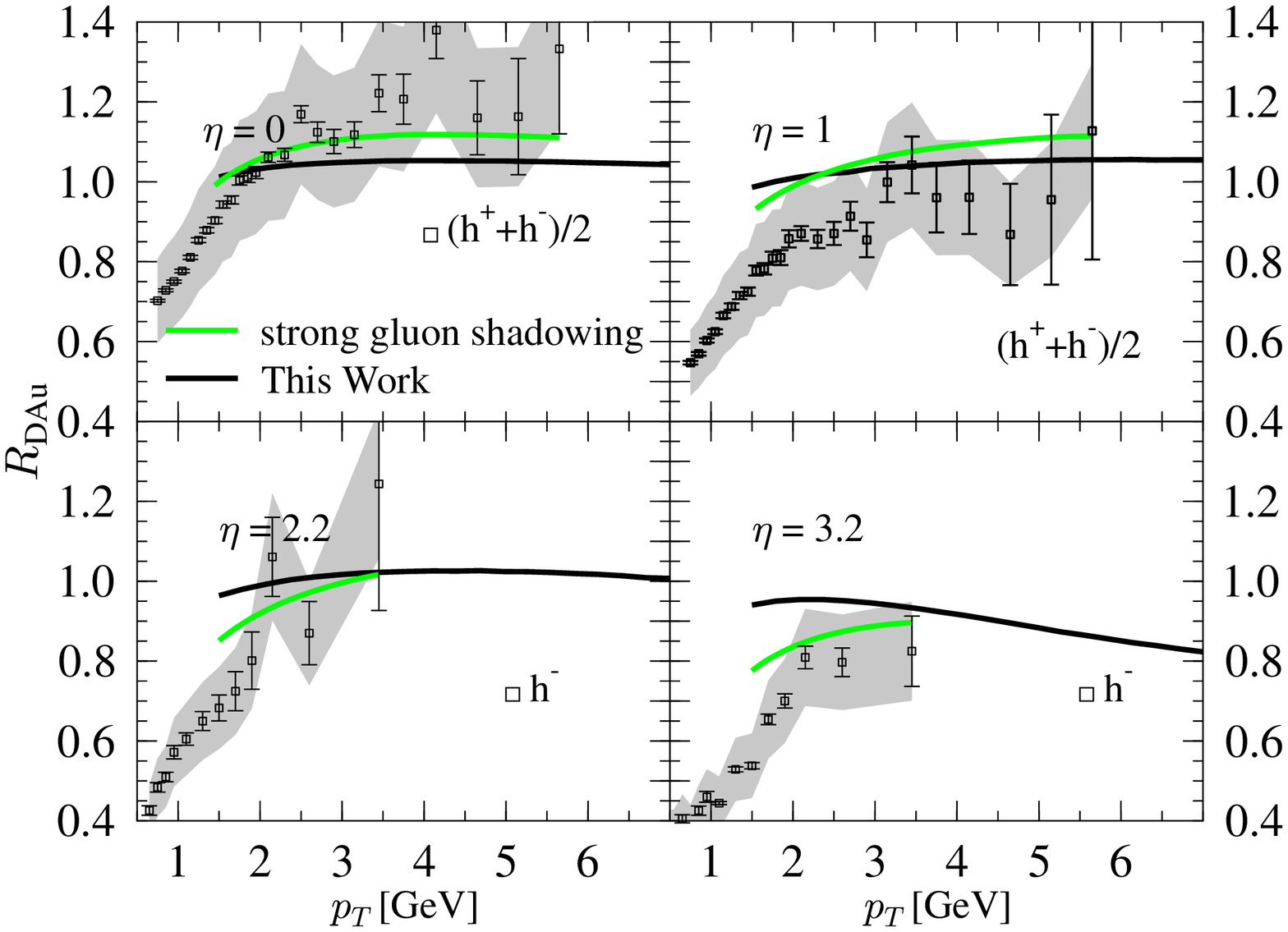}
\end{minipage}\hspace{4pc}%
\begin{minipage}{16pc}
\includegraphics[width=13pc]{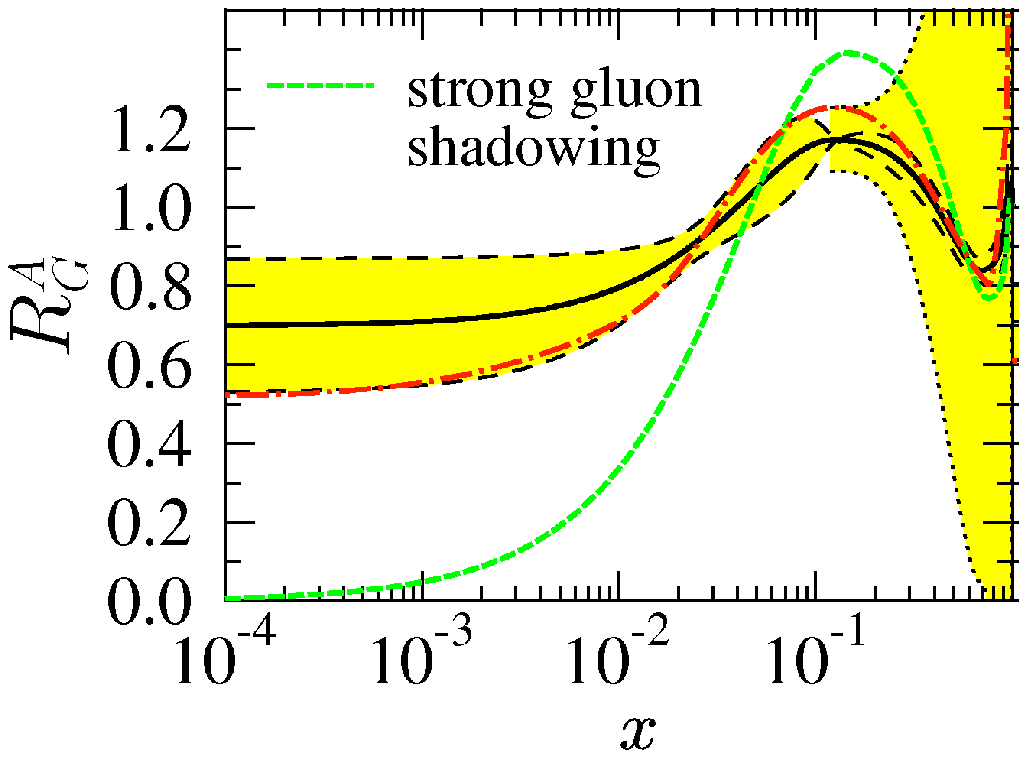}
\end{minipage} 
\caption{\label{strong}Minimum bias inclusive hadron production cross sections in D+Au collisions divided 
by that in p+p collisions at $\sqrt{s}_{NN}=200$ GeV at RHIC. The ratio  $R_{\mathrm{DAu}}$ is shown 
as a function of hadrons transverse momentum at four different pseudorapidities. 
The BRAHMS data \cite{Arsene:2004ux} are shown with the statistical error bars and the shaded 
systematic error limits.  A pQCD calculation for $h^++h^-$ production with the nuclear 
modifications from present work and KKP fragmentation functions is shown by the black lines, and that with the 
strong gluon shadowing, shown at right, by green lines.}
\end{figure}

The DY and DIS data leave the gluons very unconstrained --- gluon dependence comes only through the DGLAP-evolution. One possible way to constrain the gluon sector could be the inclusive hadron production at RHIC. Figure \ref{strong} shows data from the BRAHMS collaboration \cite{Arsene:2004ux} for $R_{\rm DAu}$, the ratio between charged hadron production in D+Au and p+p collisions as a function of hadron's $p_T$.
From the point-of-view of pQCD, this is computed via
\begin{equation}
\sigma^{AB\rightarrow h+X} = \sum_{ijkl} 
f_i^A (x_1,Q) \otimes f_j^B(x_2,Q) \otimes \sigma^{i+j\rightarrow k+l} 
\otimes D_{k\rightarrow h+X}(z,Q_f),
\label{Eq:Fragmentation}
\end{equation} 
where the $D(z,Q^2)$s are the fragmentation functions. In Fig.~\ref{strong} the pQCD calculation which employs the nPDFs from the present analysis and KKP \cite{Kniehl:2000fe} fragmentation functions, is shown by dark line. It is evident that, especially at very forward direction, our prediction is above the data and the shape is not well reproduced.

Since hadron production at the kinematical corner of low-$p_T$ and forward rapidity reaches the small-$x$ region of PDFs where the gluon distributions are the dominant ones, this could be signaling a larger uncertainty in our gluon modifications than seen in Fig.~\ref{ini}. For this reason we present an example of gluon shadowing that is much stronger. This is shown on the right hand side of Fig. \ref{strong}, and it really helps: using this gluon modification the corresponding curve for $R_{\rm DAu}$ is brought clearly closer to the BRAHMS data.

However, it's still too early to draw very strong conclusions from this observation, but a systematic study in the context of global DGLAP analysis is needed \cite{processing}. One should also bear in mind that here we are considering effects at very low-pT region, where the simple LO pQCD picture is pushed to its very limits --- although one can argue that in ratios like $R_{dAu}$ some higher order effects would partially cancel. Anyway, one should be very careful when interpreting these results. For example, it has been conjectured, that this particular BRAHMS data set could be a sign of parton saturation at work \cite{satur}.

\section*{Acknowledgements}
CAS is supported by the 6th Framework Programme of the European Community under the Marie Curie
contract MEIF-CT-2005-024624. We thank the Academy of Finland, Project 115262, and the
National Graduate School of Particle and Nuclear Physics in Finland for financial support.


\section*{References}
\begin{thebibliography}{99}

\bibitem{Arneodo:1992wf}
  M.~Arneodo,
  Phys.\ Rept.\  {\bf 240} (1994) 301.
  
\bibitem{Armesto:2006ph}
  N.~Armesto,
  J.\ Phys.\ G {\bf 32} (2006) R367
  [arXiv:hep-ph/0604108].

\bibitem{DGLAP} 
Y.~L.~Dokshitzer,
Perturbation Theory In Quantum 
Sov.\ Phys.\ JETP {\bf 46} (1977) 641
[Zh.\ Eksp.\ Teor.\ Fiz.\  {\bf 73} (1977) 1216];
V.~N.~Gribov and L.~N.~Lipatov,
Yad.\ Fiz.\  {\bf 15} (1972) 781
[Sov.\ J.\ Nucl.\ Phys.\  {\bf 15} (1972) 438];
V.~N.~Gribov and L.~N.~Lipatov,
Yad.\ Fiz.\  {\bf 15} (1972) 1218
[Sov.\ J.\ Nucl.\ Phys.\  {\bf 15} (1972) 675];
G.~Altarelli and G.~Parisi,
Nucl.\ Phys.\ B {\bf 126} (1977) 298.


  
\bibitem{Eskola:1998iy}
  K.~J.~Eskola, V.~J.~Kolhinen and P.~V.~Ruuskanen,
  Nucl.\ Phys.\ B {\bf 535} (1998) 351
  [arXiv:hep-ph/9802350].

\bibitem{Eskola:1998df}
  K.~J.~Eskola, V.~J.~Kolhinen and C.~A.~Salgado,
  Eur.\ Phys.\ J.\ C {\bf 9} (1999) 61
  [arXiv:hep-ph/9807297].

\bibitem{Hirai:2001np}
  M.~Hirai, S.~Kumano and M.~Miyama,
  Phys.\ Rev.\ D {\bf 64} (2001) 034003
  [arXiv:hep-ph/0103208].

\bibitem{Hirai:2004wq}
  M.~Hirai, S.~Kumano and T.~H.~Nagai,
  Phys.\ Rev.\ C {\bf 70} (2004) 044905
  [arXiv:hep-ph/0404093].

\bibitem{deFlorian:2003qf}
  D.~de Florian and R.~Sassot,
  Phys.\ Rev.\ D {\bf 69} (2004) 074028
  [arXiv:hep-ph/0311227].

\bibitem{Eskola:2007my}
  K.~J.~Eskola, V.~J.~Kolhinen, H.~Paukkunen and C.~A.~Salgado,
  JHEP {\bf 0705} (2007) 002
  [arXiv:hep-ph/0703104].

\bibitem{Arsene:2004ux}
  I.~Arsene {\it et al.}  [BRAHMS Collaboration],
  Phys.\ Rev.\ Lett.\  {\bf 93} (2004) 242303
  [arXiv:nucl-ex/0403005].

\bibitem{Pumplin:2002vw}
  J.~Pumplin, D.~R.~Stump, J.~Huston, H.~L.~Lai, P.~Nadolsky and W.~K.~Tung,
  JHEP {\bf 0207} (2002) 012
  [arXiv:hep-ph/0201195].

\bibitem{Hirai:2003pm}
  M.~Hirai, S.~Kumano and N.~Saito  [Asymmetry Analysis Collaboration],
  Phys.\ Rev.\  D {\bf 69} (2004) 054021
  [arXiv:hep-ph/0312112].


\bibitem{Alde:1990im}
  D.~M.~Alde {\it et al.},
  Phys.\ Rev.\ Lett.\  {\bf 64} (1990) 2479.

\bibitem{Gomez:1993ri}
  J.~Gomez {\it et al.},
  Phys.\ Rev.\ D {\bf 49} (1994) 4348.

\bibitem{Adams:1995is}
  M.~R.~Adams {\it et al.}  [E665 Collaboration],
  Z.\ Phys.\ C {\bf 67} (1995) 403
  [arXiv:hep-ex/9505006].

  \bibitem{Amaudruz:1995tq}
  P.~Amaudruz {\it et al.}  [New Muon Collaboration],
  Nucl.\ Phys.\  B {\bf 441} (1995) 3
  [arXiv:hep-ph/9503291].
  
\bibitem{Arneodo:1995cs}
  M.~Arneodo {\it et al.}  [New Muon Collaboration.],
  Nucl.\ Phys.\ B {\bf 441} (1995) 12
  [arXiv:hep-ex/9504002].



\bibitem{Arneodo:1996ru}
  M.~Arneodo {\it et al.}  [New Muon Collaboration],
  Nucl.\ Phys.\ B {\bf 481} (1996) 23.

\bibitem{Kniehl:2000fe}
  B.~A.~Kniehl, G.~Kramer and B.~Potter,
  Nucl.\ Phys.\  B {\bf 582} (2000) 514
  [arXiv:hep-ph/0010289].

\bibitem{processing}
  K.~J.~Eskola, H.~Paukkunen and C.~A.~Salgado,
  in preparation.

  \bibitem{satur}
  R.~Baier, A.~Kovner and U.~A.~Wiedemann,
  Phys.\ Rev.\ D {\bf 68}, 054009 (2003);
  D.~Kharzeev, Y.~V.~Kovchegov and K.~Tuchin,
  Phys.\ Rev.\ D {\bf 68} (2003) 094013;
  J.~L.~Albacete, N.~Armesto, A.~Kovner, C.~A.~Salgado and U.~A.~Wiedemann,
  Phys.\ Rev.\ Lett.\  {\bf 92}, 082001 (2004).

\end{thebibliography}
\end{document}